\documentstyle[prl,aps,epsfig,floats]{revtex}

\begin{document}

\wideabs{
\title{Gap generation in the XXZ model in a transverse magnetic field}
\author{D.V.Dmitriev, V.Ya.Krivnov, A.A.Ovchinnikov}
\address{Joint Institute of Chemical Physics of RAS,
Kosygin str.4, 117977, Moscow, Russia \\
and Max-Planck-Institut fur Physik Komplexer Systeme,
Nothnitzer Str. 38, 01187 Dresden, Germany.}
\maketitle
\begin{abstract}
The ground state phase diagram of the 1D XXZ model in a transverse
magnetic field is obtained. It consists of the gapped phases with
different types of long range order (LRO) and critical lines at
which the gap and the LRO vanish. Using scaling estimations and a
mean-field approach as well as numerical results we found critical
indices of the gap and the LRO in the vicinity of critical lines.
 75.10.Jm - Quantized spin models
\end{abstract}
} 

The study of the 1D spin-1/2 $XXZ$ model in a transverse magnetic field has
been drawn much attention last years. The Hamiltonian of this model is:
\begin{equation}
H=\sum (S_n^xS_{n+1}^x+S_n^yS_{n+1}^y+\Delta S_n^zS_{n+1}^z)-h\sum
S_n^x \label{H}
\end{equation}

The spectrum of the $XXZ$ model for $-1<\Delta \leq 1$ is gapless. When the
transverse magnetic field is applied a gap in the excitation spectrum seems
to open up. It is supposed \cite{kufo} that this effect can explain the
peculiarity of low temperature specific heat in $Yb_4 As_3$ \cite{YbAs}. The
magnetic properties of this compound is described by the $XXZ$ Hamiltonian
with $\Delta \approx 0.98$ and it was shown \cite{kufo} that the magnetic
field in an easy plain induces a gap in the spectrum leading to a dramatic
decrease of the linear term in the specific heat.

At $h=0$ the model (\ref{H}) is the well-known $XXZ$ model. In the
Ising-like region $\Delta >1$ the ground state of the $XXZ$ model
has the Neel long-range order (LRO) along the $Z$ axis and there
is a gap in excitation spectrum. In the region $-1<\Delta \leq 1$
system is in the so-called spin-liquid phase with a power-low
decay of correlations. Finally, for $\Delta <-1$ the ground state
is the classical ferromagnet with the gap above the ferromagnetic
state.

At $h\neq 0$ the total $S^z$ is not conserving and the model (\ref{H}) is
not integrable, except some special cases: $\Delta =1$ and
$\Delta\rightarrow\pm\infty $. In addition, there is a `classical' line
$h_{\rm cl}(\Delta)=\sqrt{2(1+\Delta )}$, where the quantum fluctuations of
$XXZ$ model are compensated by the transverse field and the exact ground
state of (\ref{H}) is a classical one \cite{classical}. The excited states
on the classical line are generally unknown (except some of them
\cite{large}), though it is assumed that the spectrum is gapped.

In the limits $\Delta\rightarrow\pm\infty $ the model (\ref{H}) reduces to
the 1D Ising model in the transverse field (ITF), for which the phase
transition occurs at $h_c=|\Delta |/2$. At this field the gap is closed and
the LRO in $Z$ direction vanishes.

It was shown \cite{J>0} that the phase transition of this type takes place
for any $\Delta >0$. One can expect also that such a transition exists for
any finite $\Delta $ at some critical value $h=h_c(\Delta )$ and there is
the transition line connecting two limiting points
$\Delta\rightarrow\pm\infty $. Besides, there are other transition lines
characterizing by vanishing of both the gap and the LRO. These lines are:
$h=0$, $|\Delta |<1$; $\Delta =1$, $h<2$; $\Delta =-1$, $h<h_c(-1)$.
However, the critical properties in the vicinity of these transition lines
are not known yet.

Thus, we expect that the phase diagram of the model (\ref{H}) (on ($\Delta$,
$h$) plane) has a form shown on Fig.1. It contains four regions
corresponding to different phases and separated by the transition lines at
which the gap vanishes. Each phase is characterized by its own type of the
LRO: the Neel order along the $Z$ axis in the region (1); the ferromagnetic
order along the $Z$ axis in the region (2); the Neel order along the $Y$
axis in the region (3); and in the region (4) there is no LRO except
magnetization along the field direction $X$ (which, certainly, exists in all
above regions).

In this paper we investigate the behavior of the gap and the LRO near the
transition (critical) lines. We are interested in the critical exponents
along these lines.

\begin{figure}[]
\vspace{0mm}
\unitlength=1cm
\begin{picture}(8,5.5)
\centerline{\psfig{file=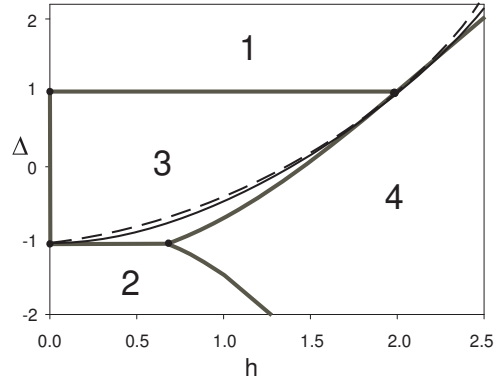,width=8cm}}
\end{picture}
\vspace{-0mm}
\caption{Phase diagram of the model (\ref{H}). The thick solid lines denote
the critical lines, thin solid line is the `classical' line, and dashed line
denotes the line $h_1(\Delta )$ (see below).}
\label{phase}
\end{figure}

{\it The line $h=0$, $|\Delta |<1$.}\nopagebreak

Low-energy properties of the $XXZ$ model are described by a free boson field
theory. Therefore, to study the behavior of the system near the line $h=0$,
$|\Delta |<1$, we use conformal estimations of small perturbation $h\ll 1$.

The time-dependent correlation functions of the $XXZ$ chain show the
power-law decay at $|\Delta |<1$ and have the asymptotic form \cite{Luther}
\begin{equation}
\langle S^x(x,t)S^x(0,0)\rangle \sim \frac{(-1)^x A_1} {\left(x^2-v^2t^2%
\right)^{\frac{\theta}{2}}} -\frac{A_2}{\left( x^2-v^2t^2\right)^{\frac{%
\theta}{2}+\frac 1{2\theta}}}  \label{XXh0}
\end{equation}
with $\theta =1-\arccos (\Delta )/\pi $, $v$ is the spin-wave velocity and
$A_1,A_2$ are constants \cite{Lukyanov}.

The non-oscillating term in Eq.(\ref{XXh0}) gives scaling dimension for
operator $S^x$ -- $d=\theta /2+1/2\theta $ and from the common formula
\cite{book} for mass gap $m$ one has:
\begin{equation}
m\sim h^\nu ,\qquad \nu =\frac 1{2-d}=\frac 2{4-\theta -1/\theta }
\label{malpha}
\end{equation}

From Eq.(\ref{malpha}) one could conclude that the magnetic field
becomes irrelevant for $\Delta<-\cos\left[\pi\sqrt{3}\right]\simeq
-0.67$ and the gap disappears for $\Delta<-0.67$. This does not
look physically reasonable, since the magnetic field destroys
continuous symmetry of $XXZ$ model and must produce the gap. In
fact, due to nonzero conformal spin $S=1$ of the non-oscillating
part of the operator $S^x$ it is necessary to consider
higher-order effects in $h$ \cite{Nersesyan}. The analysis shows,
that in the perturbation series another critical exponent appears,
giving for the mass gap
\begin{equation}
m\sim h^\gamma ,\qquad \gamma =\frac 1{1-\theta }  \label{mgamma}
\end{equation}

It turns out that the oscillating part of the operator $S^x$ gives another,
more relevant index for the gap at $\Delta<0$. Let us reproduce usual
conformal line of arguments for this oscillating part.

The perturbed action for the model is
\begin{equation}
S=S_0 + h\int {\rm d}t{\rm d}x\,S^x(x,t)  \label{action}
\end{equation}
where $S_0$ is the Gaussian action of $XXZ$ model. Let us perform an
infinitesimal renorm-group step with a scale factor $\lambda =1-\frac{\delta
L}{L}$, so that $x=\lambda x^{\prime}$, $t=\lambda t^{\prime }$. The
correlation length changes as $\xi =\lambda\xi^{\prime}$. Then, the action
becomes
\[
S^{\prime }=S_0 + h\int {\rm d}\left( \lambda t^{\prime }\right) {\rm d}%
\left( \lambda x^{\prime }\right) \,S^x(\lambda x^{\prime },\lambda
t^{\prime })
\]

Now let us estimate the large-distance contribution to the action of the
oscillating part of the operator $S^x(x,t)$:
\begin{eqnarray*}
&&h\int {\rm d}t{\rm d}x\,S^x(x,t) \sim h\int {\rm d}t\sum_n\frac{(-1)^n} {%
(n^2-v^2t^2)^{\theta /4}} \\
&& \sim  h\int {\rm d}t\sum_{n=2m}\frac{\theta
n}{(n^2-v^2t^2)^{\frac{\theta}{4}+1}} \sim h\int {\rm d}t{\rm d}x
\frac{\theta x}{(x^2-v^2t^2)^{\frac{\theta}{4}+1}}
\end{eqnarray*}

So, after rescaling we get
\[
h\int {\rm d}t{\rm d}x\,S^{x}(x,t)\rightarrow h\lambda ^{1-\theta /2}\int
{\rm d}t^{\prime }{\rm d}x^{\prime }\,S^{x}(x^{\prime },t^{\prime })
\]
and, therefore, the magnetic field scales as $h^{\prime
}=h\lambda^{1-\theta/2}$.

Expressing $\lambda $ as
\[
\lambda =\frac{\xi }{\xi ^{\prime }}=\frac{m^{\prime }}{m}=\left( \frac{%
h^{\prime }}{h}\right) ^{\frac{1}{1-\theta /2}}
\]
we find that the mass gap is proportional to
\begin{equation}
m\sim h^{\tau },\qquad \tau =\frac{1}{1-\theta /2}  \label{mbeta}
\end{equation}

Actually, the oscillating factor $(-1)^n$ in the correlator in some sense
eliminates one singular integration over $x$, and into common conformal
formula $m\sim h^{\frac 1{D-d}}$, where $D$ is the dimension of space and $d$
is the scaling dimension of perturbation operator, one should use $D=1$
instead of conventional $D=2$.

The comparison of the expressions Eqs.(\ref{malpha}), (\ref{mgamma}) and
(\ref{mbeta}) shows that for $0<\Delta<1$ the leading term is given by
Eq.(\ref{malpha}), while for $-1< \Delta <0$ by Eq.(\ref{mbeta}). Thus,
one has:
\begin{eqnarray}
m &\sim &h^\nu ,\qquad 0<\Delta <1  \nonumber \\
m &\sim &h^\tau ,\qquad -1<\Delta <0  \label{mh0}
\end{eqnarray}

For example, $m\sim h$, when $\Delta \rightarrow \pm 1$ and $m\sim h^{4/3}$
for $\Delta=0$.

In this respect the model (\ref{H}) is different from the $XXZ$ model in the
staggered transverse field for which $m\sim h^{2/(4-\theta )}$ for all
$|\Delta |<1$ \cite{Affleck}.

The staggered magnetization (LRO) along $Y$ axis behaves as:
\begin{equation}
\left\langle S_n^y\right\rangle \sim (-1)^n/\xi ^{\theta /2}\sim
(-1)^nm^{\theta /2}  \label{LRO}
\end{equation}
Therefore, the LRO has also two different critical indices:
\begin{eqnarray}
\left\langle \left| S^y\right| \right\rangle &\sim &h^{\frac \theta {%
4-\theta -1/\theta }}\qquad 0<\Delta <1  \nonumber \\
\left\langle \left| S^y\right| \right\rangle &\sim &h^{\frac \theta {%
2-\theta }}\qquad -1<\Delta <0  \label{LROh0}
\end{eqnarray}

The critical indices $\nu$ and $\tau$ can be also found from the
analysis of divergences of terms of perturbation series in $h$ at
$N\rightarrow\infty$ ($N$ is the system size). As it is shown
\cite{large} at $N\rightarrow\infty$ the ground state energy has a
form
\begin{equation}
\frac{\delta E}N =-\frac \chi 2h^2+ah^{2\nu }+bh^{2\tau }  \label{dE}
\end{equation}
where $\nu$ and $\tau$ are given by Eqs.(\ref{malpha}) and (\ref{mbeta}) and
$a,b$ are some constants.

One can see from Eqs.(\ref{malpha},\ref{mbeta}) that $\nu \rightarrow 1$ at
$\Delta \rightarrow 1$ and $\tau \rightarrow 1$ at $\Delta \rightarrow -1$.
Hence, in both limits one of the singular terms becomes proportional to
$h^2$, and, therefore, gives a contribution to the susceptibility.
It implies that in the symmetric points $\Delta =\pm 1$ the susceptibility
has a jump. For example, $\chi =1/4$ at $\Delta =-1$ and $\chi =1/8$ at
$\Delta\rightarrow -1$ \cite{large}.

{\it The line $\Delta =1$.}\nopagebreak

In the vicinity of the line $\Delta =1$ it is convenient to rewrite the
Hamiltonian (\ref{H}) in the form
\begin{eqnarray}
H &=&H_0+V  \nonumber \\
H_0 &=&\sum {\bf S}_n\cdot {\bf S}_{n+1}-h\sum S_n^x  \nonumber \\
V &=&-g\sum S_n^zS_{n+1}^z  \label{Hd1}
\end{eqnarray}
where the parameter $g=(1-\Delta )\ll 1$ is small. On the isotropic line
$\Delta =1$ the model (\ref{H}) is exactly solvable by Bethe ansatz. The
ground state of $H_0$ remains spin-liquid one up to the transition point
$h_c=2$, where the phase transition of the Pokrovsky-Talapov type takes
place and the ground state becomes completely ordered ferromagnetic state.
Therefore, for $h<2$ and for small perturbation $V$ we can use conformal
estimations.

The large distance asymptotic of the correlation function on this line is
\begin{equation}
\langle S_i^zS_{i+n}^z\rangle \sim \frac{(-1)^n}{n^{\alpha (h)}},
\label{ZZd1}
\end{equation}
where $\alpha (h)$ is known function obtained from Bethe ansatz \cite{BIK}
and having the following limits:
\begin{equation}
\alpha (h) \sim 1-\frac 1{2\ln \left( 1/h\right) }\qquad h\rightarrow 0
\label{alphah0}
\end{equation}
and $\alpha(2)=1/2$.

So, the scaling dimension of operator $S^z$ is $d_z=\alpha(h)/2$, and the
scaling dimension of operator $S_i^zS_{i+1}^z$ is $d_{zz}=4d_z=2\alpha (h)$.
Since $\alpha (h)<1$, then the perturbation $V$ is relevant and leads to the
mass gap and the staggered magnetization along $Y$ axis
\begin{eqnarray}
m &\sim &g^{1/(2-d_{zz})}=g^{1/(2-2\alpha )}  \label{md1} \\
\left\langle \left| S^y\right| \right\rangle &\sim &1/\xi ^{d_z}\sim
g^{\alpha /(4-4\alpha )}  \label{LROd1}
\end{eqnarray}

Above consideration is valid also for the case $\Delta>1$ (region 1) with
the same exponents for the gap and the LRO. The only difference is that the
staggered magnetization appears in this case along the $Z$ axis.

From the general expressions for the mass gap (\ref{malpha}), in the limit
$h\rightarrow 2$ we obtain that $m\sim g$.

The LRO in the vicinity of the point $\Delta =1,h=2$ vanishes on both lines:
at $\Delta =1$ from (\ref{md1}) as $g^{1/4}$; and at $h=h_c$ as $\left|
h_c-h\right| ^{1/8}$ (see Eq.(\ref{eqfermi})). Combining these facts we
arrive at the following formula:
\begin{equation}
\left\langle S_n^y\right\rangle \simeq (-1)^ng^{1/4}\left| h_c-h\right|
^{1/8}  \label{LROd1h2}
\end{equation}
which is in accordance with the exact expression for LRO on the classical
line \cite{classical}.

The behavior of the system near the point $\Delta =1$, $h=0$ is more
complicated. As it follows from Eq.(\ref{mh0}), for very small $h$ the mass
gap is $m\sim h$, while on the other hand from Eq.(\ref{md1}) one obtains
another scaling $m\sim g^{\ln(1/h)}$. Really \cite{Schulz,large}, there are
two regions near this point with different behavior of the mass gap and a
crossover line $\sqrt{g}\ln(1/h)\sim 1$:
\begin{eqnarray}
m &\sim &h,\qquad\qquad{\rm for\quad}\sqrt{g}\ln(1/h)\gg 1  \nonumber \\
m &\sim &g^{\ln(1/h)},\qquad{\rm for\quad}\sqrt{g}\ln(1/h)\ll 1
\label{md1h0}
\end{eqnarray}

{\it The transition line $h=h_c(\Delta )$.}\nopagebreak

Now we consider the behavior of the model in the vicinity of the transition
line $h_c(\Delta )$. For this we have used the Fermi-representation of
(\ref{H}). At first, in (\ref{H}) we perform a rotation of the spins around
the $Y$ axis by $\pi/2$ (so that the magnetic field will be directed along
the $Z$ axis) followed by the Jordan-Wigner transformation to the Fermi
operators $a_n^+,a_n$. As a result, we obtain the Fermi Hamiltonian in
the form
\begin{eqnarray}
H_{\rm F} &=&-\frac{hN}2+\frac N4+\sum (h-1-\frac{1+\Delta }2\cos
k)a_k^{+}a_k  \nonumber \\
&+& \frac{1-\Delta }4\sum \sin k(a_k^{+}a_{-k}^{+}+a_{-k}a_k)  \nonumber \\
&+& \sum a_n^{+}a_na_{n+1}^{+}a_{n+1}  \label{Hfermi}
\end{eqnarray}

Treating the Hamiltonian $H_{\rm F}$ in the mean-field approximation we
find the ground state energy $E_0$ and the excitation spectrum
$\varepsilon(k)$.

Main results following from the mean-field consideration are:

\noindent 1. The function $\varepsilon (k)$ has a minimum at
$k_{\rm min}$, which is changed from $\pi/2$ at $h=0$ to zero at
$h=h_1(\Delta)$ and $k_{\rm min}=0$ for $h>h_1(\Delta)$. The gap
in the spectrum $\varepsilon (k)$ vanishes at $h_c(\Delta )$
($h_c>h_1$) and for $h>h_1$ is $m\sim |h-h_c|$. The dependencies
of $h_1(\Delta )$ and $h_c(\Delta )$ are shown on Fig.1. There is
the staggered magnetization along $Y(Z)$ axis for $\Delta <1
(\Delta
>1)$ at $h<h_c$ and it behaves as $\sim |h-h_c|^{1/8}$ at
$h\rightarrow h_c$. The magnetization $s=\left\langle
S_n^x\right\rangle$ has a logarithmic singularity at $h\rightarrow
h_c$.

These results show that the transition at $h=h_c(\Delta )$ belongs to the
universality class of the ITF model.

\noindent 2. The mean field approximation is exact on the classical line
$h=h_{\rm cl}(\Delta )$.

\noindent 3. In the vicinity of the point $h=2$, $\Delta =1$ the fermion
density is small and the mean field approximation gives the accuracy in the
energy, at least, up to $g^3$ or $(2-h)^4$. For this case the gap is
\begin{eqnarray}
m &=&|h-h_c|,\qquad h>h_1  \nonumber \\
m &=&\frac{g}{2\sqrt{2}}\sqrt{h_c-h-\frac{g^2}{32}},\qquad h<h_1
\label{eqfermi}
\end{eqnarray}
where $h_c=2-\frac g2 -\frac{g^2}{32}$, $h_1=h_c-\frac{g^2}{16}$.

It is interesting to compare Eqs.(\ref{eqfermi}) with the conformal
estimation of the gap $m\sim g$. The conformal approach determines $g$
dependence only, while Eq.(\ref{eqfermi}) gives a prefactor depending on $h$.

The magnetic susceptibility $\chi (h)$ is
\begin{eqnarray}
\chi &=&\frac 2{\pi g}\ln \left( \frac{g^2}{h_c-h}\right) ,\qquad g\gg \sqrt{%
h_c-h}  \nonumber \\
\chi &=&\frac 1{\sqrt{2}\pi }\frac 1{\sqrt{h_c-h}},\qquad g\ll \sqrt{h_c-h}
\label{susfermi}
\end{eqnarray}

As it follows from (\ref{susfermi}) there is a crossover from square root to
logarithmic divergence of $\chi $.

{\it The line $\Delta =-1$.}\nopagebreak

On the line $\Delta =-1$ the model (\ref{H}) reduces to the isotropic
ferromagnet in a staggered magnetic field. This model is non-integrable, but
it was suggested \cite{Alkaraz}, that the system is governed by a $c=1$
conformal field theory up to some critical value $h=h_0$, where the phase
transition of the Kosterlitz-Thouless type takes place. In the vicinity of
the line $\Delta =-1$ the Hamiltonian (\ref{H}) becomes
\begin{eqnarray}
H&=&-\sum {\bf S}_n\cdot {\bf S}_{n+1}-h\sum (-1)^nS_n^x \nonumber \\
&+&(1+\Delta )\sum S_n^zS_{n+1}^z  \label{Hd-1}
\end{eqnarray}
where $(1+\Delta )\ll 1$ is a small parameter.

It can be shown \cite{large} that at $h_{\rm cl}(\Delta )<h\ll 1$
low energy states of the (\ref{Hd-1}) is described by the $XYZ$
Hamiltonian
\begin{equation}
H=-\sum [(1-\frac{h^2}{2})S_n^x S_{n+1}^x + S_n^y S_{n+1}^y -\Delta S_n^z
S_{n+1}^z]  \label{HXYZ}
\end{equation}

The derivation of this mapping is based on the fact that the
transition operator $\sum (-1)^n S_n^x$ connects the low-lying
states with the states with high energies $\sim 2$ only. The
coincidence of the low-energy spectra of (\ref{Hd-1}) and
(\ref{HXYZ}) for $h_{\rm cl}(\Delta )<h\ll 1$ has been checked
numerically. The spectrum of low-lying excitations of the $XYZ$
model \cite{kosevich} as well as initial model (\ref{H}) in the
vicinity of the point $\Delta=-1,h=0$ can be described
asymptotically exactly by the spin-wave theory \cite{kufo}, which
gives
\begin{eqnarray}
m &=&h\sqrt{(1+\Delta )/2},{\rm \qquad }\Delta >-1  \nonumber \\
m &=&\sqrt{(1+\Delta )(1+\Delta -h^2/2)},{\rm \qquad }\Delta <-1
\label{mXYZ}
\end{eqnarray}

The validity of the spin-wave approximation is quite natural because in the
vicinity of the point $\Delta=-1,h=0$ the number of magnons forming the
ground state is small.

We note that the gap (\ref{mXYZ}) for $\Delta\geq -1$ coincides with the
conformal theory result (\ref{mh0}) and provides us with preexponential
factor for the gap.

On the line $\Delta =-1$ the model (\ref{HXYZ}) is the $XXZ$ model
and the correlation functions have the power-law decay. The
scaling dimensions of operators $S_i^x$ and $S_i^y,S_i^z$ on this
line are $d_x=\beta(h)/2$ and $d_y=d_z=1/2\beta(h)$. The function
$\beta(h)$ is generally unknown, but at $h\ll 1$, where the
mapping to $XYZ$ model is valid, $\beta(h)\sim\pi/h$.

Strictly on the line $\Delta =-1$ at some value of $h=h_0$ the gap
appears. It means that the magnetic field term is irrelevant at
$h<h_0$ ($\beta (h)>4$) and becomes marginal at $h=h_0$, where
$\beta (h_0)=4$. So, at the point $h=h_0$ the transition is of the
Kosterlitz-Thouless type, and for $h>h_0$ the mass gap is
exponentially small.

Using the conformal invariance of the model (\ref{Hd-1}) at
$\Delta =-1$ and $h<h_0$ we carried out finite-size calculation of
the exponent $\beta(h)$. The extrapolated function $\beta(h)$
agree well with the dependence $\pi/h$ at $h\ll 1$ and $\beta=4$
at $h_0\simeq 0.52$.
On the other hand, the mean-field approach gives rather crude
value $h_0=h_c(-1)=0.69$.

In summary, we have studied the 1D XXZ model in the transverse magnetic
field. It is shown the spectrum of the model is gapped except critical lines
on the $(h,\Delta)$ plane, where the LRO vanishes. We found the critical
exponents of the gap and the LRO in the vicinity of these lines.

Authors would like to thank Prof. P.Fulde and Dr. A.Langari for
helpful discussions. Authors are grateful to Max-Planck-Institut
fur Physik Komplexer Systeme for a kind hospitality. This work is
supported under RFFR Grants No 00-03-32981 and No 00-15-97334.


\end{document}